\documentclass[aps,twocolumn,longbibliography,pra,superscriptaddress,amssymb,amsmath,amsmath,floatfix]{revtex4-1}
\usepackage{graphicx}
\usepackage{epstopdf}
\usepackage{comment}
\usepackage{bm}
\usepackage{wrapfig}
\usepackage{times}
\usepackage[colorlinks=true,linkcolor=blue,urlcolor=blue,citecolor=blue]{hyperref}
\usepackage{float}
\usepackage{titlesec}

\begin{document}

\title{Ground and excited potential energy surfaces for CaF+Ca interactions and isotope exchange reactions}

\author{Dibyendu Sardar}
\email{chem.dibyandu.sardar@gmail.com}
\affiliation{Faculty of Physics, University of Warsaw, Poland}

\author{John L. Bohn}
\affiliation{JILA, NIST, and Department of Physics, University of Colorado, Boulder, Colorado 80309, USA}

\date{\today}
\begin{abstract}
We investigate the intermolecular interactions between laser-cooled CaF and Ca, in their ground and excited electronic states, aiming to understand atom-exchange reaction pathways. Using state-of-the-art \textit{ab initio} quantum chemistry methods, we compute potential energy surfaces for nine electronic states arising from the lowest three asymptotes of Ca$_2$F trimer, within the rigid rotor approximation applied to CaF. Two-dimensional potential energy surfaces are computed for the ground state and one of the excited states. We use a combination of the coupled cluster method restricted to single, double, and perturbative triple excitations, and the multireference configuration interaction method with single and double excitations. The ground (X)~$^2\mathrm{A}'$ electronic state of the trimer is significantly deep and highly anisotropic. The excited electronic states are also strongly bound. Notably, the potential energy surface of one of the excited states, (2)~$^2\mathrm{A}'$, lies below the ground-state asymptote of the trimer. By analyzing the potential energy surfaces, we discuss atom-exchange reaction pathways involving both the ground-state interaction between CaF and Ca and the excited metastable state of Ca.

\end{abstract}

\maketitle
\section{Introduction}
\label{sec:intro}
Ultracold molecules have emerged as an impactful platform for probing atomic, molecular, and optical physics (AMO) beyond the current state of knowledge. The rich internal structure and permanent electric dipole moment make them an ideal testbed for exploring long-range physics and short-range chemistry in a new energy regime. Ultracold molecules are synthesized by combining a pair of precooled atoms through magneto-association, followed by an optical stabilization using stimulated Raman adiabatic passage \cite{VogesPRL2020,YangPRL2020,GuoPRL2016,ParkPRL2015,NiScience2008,MolonyPRL2014,StevensonPRL2023}. Another important method, direct laser cooling \cite{ShumanNAT2010,AndereggNATPHYS2018,CheukPRL2018,CaldwellPRL2019,DingPRX2020}, relies on repeated photon cycling enabled by highly diagonal Franck-Condon factors. Ultracold molecules have a range of applications from quantum computations \cite{DemillePRA2002,YelinPRA2006}, quantum simulations for many-body physics \cite{CornishNatPhys2024,MicheliNATPhys2006,BaranovCHEMREV2012}, novel quantum phases, to precision measurements for fundamental physical constants \cite{CarrNJP2009,DemilleScience2017}. Furthermore, ultracold molecules open up a new platform for exploring ultracold chemistry \cite{LiuNAT2021} with an unprecedented level of control.

The study of chemical reactions at ultracold temperatures advances the field of ultracold quantum gases \cite{KarmanNP24}. At these temperatures, chemical reactions could be surprisingly efficient, attributed to the wave nature of reactants. Reaction rates are completely guided by non-classical effects such as wave function delocalization and quantum mechanical tunneling through a barrier. Chemical reactivity was first observed in a sample of ultracold KRb molecules \cite{OspelkausScience2010}. By combining AMO physics and physical chemistry techniques, ultracold chemical reactions have been successfully followed from their initiation to completion \cite{HuScience2019}. Recently, product state distribution has been measured in ultracold chemical reactions \cite{LiuNature2021}. Ultracold chemical reactions have been controlled by engineering long-range dipolar interactions through external fields \cite{AndereggScience2021,KarmanPRL2018,QuemenerPRA2011}. 

Theoretically, ultracold chemical reactions have been investigated in alkali-metal and alkaline-earth dimers. For the singlet ground-state \cite{ZuchowskiPRA2010} of alkali dimers, atom-exchange reactions to dimer formation are energetically allowed for some molecules, while atom-transfer reactions to trimer formation are energetically forbidden. In contrast, atom-exchange and atom-transfer reactions are energetically allowed for the first excited triplet state \cite{TomzaPRA2013} of alkali dimers. Again, in the ground state of alkaline-earth molecules \cite{LadjimiPRA2023}, both the dimer- and trimer-forming reactions are exothermic. An exothermic chemical reaction has been predicted for the ground-state RbSr molecule \cite{ManPRA2022}. For all these cases, the allowed chemical reactions proceed through a submerged barrier. Furthermore, chemical reactions are energetically allowed for the ground state of CaF \cite{SardarPRA2023} and SrF \cite{MeyerPRA2011} dimers. In parallel, exothermic isotope-exchange chemical reactions have been studied between ground-state alkali-metal and alkaline-earth dimers \cite{TomzaPRL2015}. Barrierless isotope exchange reaction and product state distribution have also been investigated in the ultracold SrF+Sr system \cite{KosickiJPCA2017,KosickiJPB2020}. 

\begin{table*}
\caption{\label{tab:channel} Molecular states for Ca$_2$F trimer arising from the lowest four asymptotes under the C$_\mathrm{s}$ point group. $\Delta E$ (in cm$^{-1}$) is the difference in energy with respect to the ground state asymptote of CaF+Ca. }
\begin{ruledtabular}
\begin{tabular}{ccc}
Asymptote  & Molecular states CaF+Ca&$\Delta E$ (cm$^{-1}$)\\
\hline
 CaF$(^2\Sigma^+)$ + Ca$(^{1}\mathrm{S})$ &    $^2\mathrm{A}'$& 0 \\ 
 CaF$(^2\Sigma^+)$ + Ca$(^{3}\mathrm{P})$ & $^2\mathrm{A}'$ ($\times 2$), $^2\mathrm{A}''$, $^4\mathrm{A}'$ ($\times 2$), $^4\mathrm{A}''$&15315  \\ 
CaF$(^2\Pi)$ + Ca$(^{1}\mathrm{S})$ & $^2\mathrm{A}'$, $^2\mathrm{A}''$ &  16490 \\ 
CaF$(^2\Sigma^+)$ + Ca($^{3}\mathrm{D})$ &$^2\mathrm{A}'$ ($\times 3$), $^2\mathrm{A}''$ ($\times 2$), $^4\mathrm{A}'$ ($\times 3$), $^4\mathrm{A}''$ ($\times 2$)  &  20371 \\ 
\end{tabular}
\end{ruledtabular}
\end{table*}

Understanding ultracold chemical reactions and their underlying mechanisms requires detailed insight into the intermolecular interactions between the reactants, in both their ground and excited electronic states. Ultracold molecule-molecule interactions and their potential energy surfaces (PESs) were reported for a few alkali dimers, particularly in their ground state \cite{YangJPCL2020,LiuJCPA2022}, and excited states \cite{ChristianenJCP2019}. Ground and excited PESs for ultracold CaF+CaF have also been studied \cite{SardarPRA2023,SardarJPCA2023,SardararXiv2025}. Additionally, PESs for alkali-metal molecule-atom systems \cite{LiuNatChem2025,KarmanPRA2023} and alkaline-earth fluoride molecule with alkali atom have been studied in their ground state \cite{KosickiJPCA2017,DaPCCP2023}, including SrF+Sr, in the context of ultracold chemical reactions.

In this work, we investigate ultracold molecule-atom interactions in their ground and excited electronic states, considering CaF+Ca as a prototype system. Here, both CaF and Ca are laser-coolable species. The first excited $^3$P state of Ca is a metastable state with a lifetime of approximately 0.39 milliseconds \cite{FurcinittiPLA1975}. Thus, interactions and ultracold collisions in CaF+Ca will allow for the investigation of excited-state dynamics involving ultracold molecules and atoms. We focus on the ultracold CaF molecule, which can be trapped in magnetic and optical traps and cooled to temperatures as low as a few microkelvins \cite{AndereggNATPHYS2018,CheukPRL2018}. Besides, ultracold collisions between CaF molecules have been experimentally studied in
optical tweezer traps \cite{CheukPRL2020}, and their interactions have been theoretically explored \cite{SardarPRA2023,SardarJPCA2023}.

The primary objective of this study is to compute PESs for the ground and excited states of CaF+Ca. Next, by exploiting PESs, we investigate prospects for isotope-exchange chemical reactions involving ground-state CaF and Ca in both its ground $\mathrm{^1S}$ and metastable excited $^3\mathrm{P}$ state. These reactions could be experimentally studied and may provide an intriguing platform for exploring chemical reactions at ultracold temperatures. To date, interactions and isotope-exchange chemical reactions in the CaF+Ca system have remained unexplored in both theory and experiments. Further, knowledge of the ground and excited states could contribute to understanding of photoassociation experiments that are useful for spectroscopy and for creating trimers from the atom-dimer system.

Using state-of-the-art \textit{ab initio} quantum chemistry methods, we compute PESs for the CaF+Ca system in both its ground and excited electronic states. We compute nine molecular electronic states arising from the lowest three asymptotes of Ca$_2$F, while additional excited states that correlate with higher asymptotes are difficult to converge. In the present work, we employ an optimal active space with a quadruple-zeta basis set to describe the excited electronic states of the Ca$_2$F trimer. The ground state, X~$^2\mathrm{A}'$, of the Ca$_2$F trimer is symmetric and strongly bound. The two-dimensional variation for the ground-state PES exhibits a highly anisotropic potential. We find that ground-state interactions between CaF and Ca will allow an isotope-exchange chemical reaction that proceeds without a barrier. The excited states of the Ca$_2$F trimer are also strongly bound. 
In the excited electronic manifold, particularly when Ca is in the metastable $^3\mathrm{P}$ state, one of the PESs for the excited state, (2)~$^2\mathrm{A}'$, correlating to the CaF($^2\Sigma$)+Ca($^3\mathrm{P}$) asymptote, crosses the ground-state asymptote. We further verify this distinct feature in the PES of  (2)~$^2\mathrm{A}'$ state, by allowing relaxation of the CaF bond vibration. Notably, for a particular configuration of Ca$_2$F trimer, two-dimensional variation for the (2)~$^2\mathrm{A}'$ PES lies more than 1000 cm$^{-1}$ below the ground-state asymptote. In general, our findings may guide spectroscopic measurements for Ca$_2$F trimer and facilitate investigations of ultracold chemistry in the CaF+Ca molecule-atom system.

This paper is organized as follows. In Sec.~\ref{sec:methodology}, we describe the \textit{ab initio} methods used to construct the potential energy surfaces for CaF+Ca. In Sec.~\ref{sec:results}, we discuss one-dimensional cut and two-dimensional properties of the molecular potential energy surfaces for Ca$_2$F trimer, including the chemical reaction pathways. Finally, in Sec.~\ref{sec:conclusion}, we provide a summary and outlook of this work.
\section{Computational details}
\label{sec:methodology}

We use state-of-the-art \textit{ab initio} quantum chemical calculations to compute PES for the ground and excited electronic states of Ca$_2$F trimer. \textit{Ab initio} calculations are performed by \textsc{Molpro} 2022.1 \cite{WernerJCP2020} software package.  

The interactions of ground-state CaF and Ca result in X~$^2\mathrm{A}'$ state under C$_\mathrm{s}$ symmetry.  In the highly symmetric linear configurations of  Ca$_2$F trimer, X~$^2\mathrm{A}'$ state corresponds to the X~$^2\Sigma$ under the C$_{2{\mathrm{v}}}$ point group. However, interactions between the ground state of CaF and the metastable $^3\mathrm{P}$ state of Ca lead to the states $\mathrm{A}'$ and $\mathrm{A}''$, with doublet and quartet spin multiplicities. Furthermore, both $\mathrm{A}'$ and $\mathrm{A}''$ states with only doublet spin symmetry are obtained from the interactions of CaF($^2$P) and Ca($^1\mathrm{S}$). Table~\ref{tab:channel} presents these molecular states, corresponding to the lowest three asymptotes, as well as the states of the next asymptote of the Ca$_2$F trimer, together with the energy difference relative to the ground-state asymptote. In this study, we restrict our calculation to the lowest three asymptotes, since including additional CaF($^2\Sigma^+$)+Ca($^3\mathrm{D}$) asymptote will make the calculation highly demanding in terms of computational resources.

\begin{figure}[t]
  \includegraphics[width=0.35\textwidth]{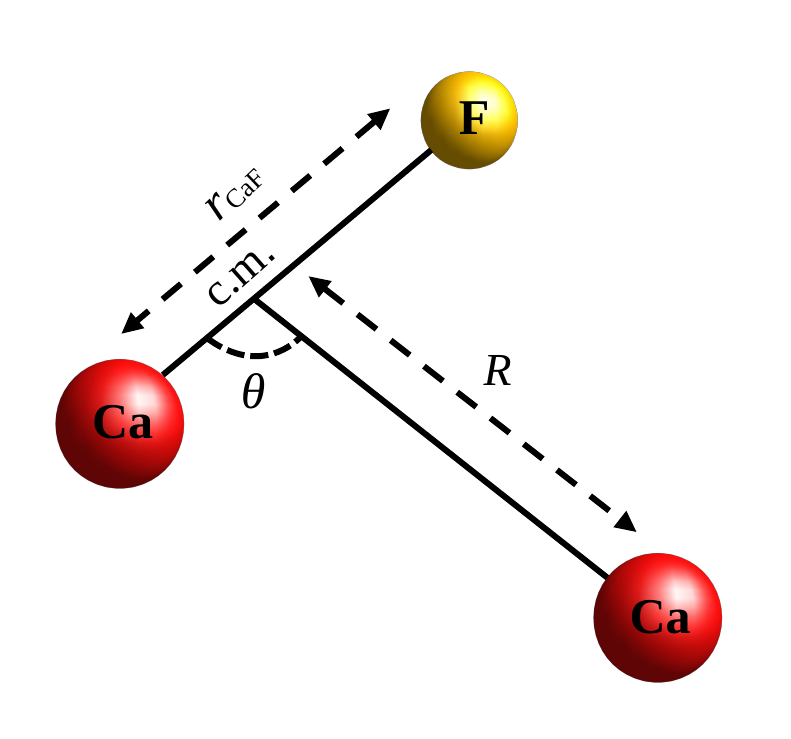}
  \caption{A schematic diagram of the CaF+Ca system in the Jacobi coordinates. 
  }\label{fig:schematic}
\end{figure}

We present triatomic Ca$_2$F in the Jacobi coordinates to compute one-dimensional (1D) cuts and two-dimensional (2D) PES. In these calculations, we consider CaF as a rigid rotor, and its bond length remains fixed at the equilibrium value ($r_{\textrm{CaF}} = 3.695$ bohr) measured experimentally \cite{KaledinJMSP1999}. In Figure~\ref{fig:schematic}, we show a schematic plot of the Jacobi coordinates for CaF+Ca, where $R$ is the distance from the center of mass (c.m.) of CaF to Ca, and $\theta$ is the angle between the molecular axis and c.m. to Ca. Under the rigid rotor approximation for CaF, the PES $V(R,\theta)$ for Ca$_2$F can be expressed in terms of Legendre polynomials $P_\lambda(^.)$
\begin{equation}
    V(R,\theta) = \sum_{\lambda=0}^{\lambda_{\text{max}}} V_\lambda (R) P_\lambda(\cos\theta). \label{eq:pes}
\end{equation}
Here, $\lambda$ denotes the order of the polynomials. $V_\lambda (R)$ represents the Legendre components of the 2D PES, and provides insight into the anisotropy of the PES. This decomposition is useful to study molecule-atom scattering dynamics through coupled-channel calculations. 

The long-range part of the potential for the ground state interaction between CaF and Ca is given by
\begin{equation}
    V(R,\theta) = -\Bigl(\frac{C^\text{ind}_{6,0}}{R^6}+\frac{C^\text{disp}_{6,0}}{R^6}\Bigr)- \Bigl( \frac{C^\text{ind}_{6,2}}{R^6}+\frac{C^\text{dis}_{6,2}}{R^6}\Bigr)P_2(\cos\theta) + ...
    \label{eq:lr}
\end{equation}
The induction component is given by
\begin{equation}
    C^\text{ind}_{6,0}=\alpha_{\textrm{Ca}}d^2_{\textrm{CaF}},
    \label{eq:ind}
\end{equation}
where $\alpha_{\textrm{Ca}}$ and $d_{\textrm{CaF}}$ are the static polarizability of Ca and permanent electric dipole moment of CaF. The dispersion components are given by
\begin{equation}
\begin{split}
  C^\text{disp}_{6,0} &= \frac{3}{\pi}\int_0^\infty \alpha_\text{Ca}(i\omega) \bar{\alpha}_\text{CaF}(i\omega) d\omega, \\
  C^\text{disp}_{6,2} &= \frac{1}{\pi}\int_0^\infty \alpha_\text{Ca}(i\omega){\Delta\alpha}_\text{CaF}(i\omega)d\omega.
  \label{eq:disp}
\end{split}  
\end{equation}
Here, $\alpha_\text{Ca}(i\omega)$ denotes the dynamic polarizability of Ca at an imaginary frequency $\omega$. The average ($\bar{\alpha}_\text{CaF}$) and anisotropic (${\Delta\alpha}_\text{CaF}$) polarizabilities are calculated from the parallel ($\alpha_{||}\equiv\alpha_{zz}$) and perpendicular ($\alpha_\perp\equiv\alpha_{xx}\equiv\alpha_{yy}$) components of the dynamic polarizability tensor of CaF and expressed as:
\begin{equation}
  \bar\alpha_\text{CaF}= (\alpha_{||}+2\alpha_\perp) /3, \hspace{0.2cm}\textrm{and}\hspace{1.0cm} {\Delta\alpha}_\text{CaF}=(\alpha_{||}-\alpha_\perp),
\end{equation}
where we consider the z-axis to be the internuclear axis of CaF.


We compute the molecular PES of Ca$_2$F trimer using a pseudopotential-based correlation-consistent basis set with diffuse augmenting functions. In particular, we consider aug-cc-pwCVQZ-PP \cite{HillJCP2017} for Ca and aug-cc-pwCVQZ for F \cite{HillJCP2017}. Scalar relativistic effects are accounted for by using a small core energy consistent pseudopotential ECP10MDF \cite{LimJCP2006} for the ten inner core electrons of Ca.

For the ground X~$^2\mathrm{A}'$ state, we consider the state-of-the-art coupled cluster method, with single, double, and perturbative triple excitations CCSD(T). Initially, we construct Hartree-Fock orbitals on grids of $R$ and $\theta$, then perform CCSD(T) calculations. We consider 44 points of $R$ ranging from 3.5 to 50 bohr, and 11 grid points in $\theta$ based on quadrature points for Legendre polynomials of order $\lambda_{\mathrm{max}}=9$, including 0 and 180 degrees. We compute the interaction energy for the trimer using the supermolecular approach, and we apply the Boys-Bernardi counterpoise correction for the basis-set superposition error.


We compute the excited electronic states of the trimer using an internally contracted multireference configuration interaction (MRCI) method. An accurate description of the excited electronic states of Ca$_2$F timer is challenging and depends on the active space (AS) concerned. Our initial calculations show that a reasonably large AS with a quadrupole-zeta basis is computationally demanding. To address this, we construct an optimal AS consisting of the highest occupied molecular orbital and the lowest four unoccupied molecular orbitals of CaF, and the 4s, 4p, and 3d atomic orbitals of Ca. Initially, we conduct a complete active space self-consistent field (CASSCF) calculation on this AS. We treat the 3s and 3p orbitals of Ca and the 1s orbital of F as closed-shell orbitals and are fully optimized at the CASSCF level. Finally, we carry out the MRCI calculation where the 1s orbital electrons of F remain uncorrelated. 

We determine the leading long-range induction and dispersion coefficients for CaF+Ca. To this end, we compute the dynamic polarizability of CaF at imaginary frequency by solving the damped-response equations at the Hartree-Fock level using  \textsc{Dalton}.2020. software package. In addition, we calculate static polarizability of CaF at the CCSD(T) level. Finally, we scale each component of the dynamic polarizability to match the static polarizability obtained from the CCSD(T) calculation. Likewise, we scale the literature-reported dynamic polarizability of Ca \cite{DereviankoADNDT2010} to reproduce its static polarizability value \cite{SchwerdtfegerMP2019}.

\section{Results and discussions}
\label{sec:results}
\subsection{Equilibrium properties of dimer and trimer}
\label{subsec:Eqprop}
The molecular radical CaF exhibits significant ionic bonding character. In the ground X~$^2\Sigma$ state, Ca transfers one of the valence electrons out of two $4s^2$ electrons to the F atom, resulting in stronger bonding. We compute the depth of the potential at the CCSD(T) and MRCI levels are 44254 cm$^{-1}$ and 43408 cm$^{-1}$, compared to the reported value 44203 cm$^{-1}$ \cite{HouJQSRT2018,CharronJMS1995}. For the first excited A~$^2\Pi$ state of CaF, we obtain the potential well depth 27481 cm$^{-1}$ and 27428 cm$^{-1}$ using the CCSD(T) and MRCI methods. For both states of CaF, the calculated parameters obtained by CCSD(T) and MRCI methods agree well with literature-reported values \cite{SardararXiv2025}.  

We perform full-dimensional geometry optimization for the ground state of Ca$_2$F trimer using the CCSD(T) method. We note that the ground X$~^2\mathrm{A}'$ state optimizes to a bent geometry. The CaF bond length elongates by approximately 10\% compared to the rigid rotor condition. The optimized parameters are tabulated in Table~\ref{tab:global}, and are comparable to the reported values \cite{KosickiJPCA2017}. Notably, the potential well depth for the Ca$_2$F trimer is about 20\% shallower than that of the CaF molecule. It indicates that the formation of a Ca$_2$F trimer requires the redistribution of electron density between two Ca metal centers and the F ligand, resulting in an elongated bond length and weaker binding in the trimer. The CaF+Ca potential well is about 50\% shallower than that of the singlet ground-state for CaF+CaF \cite{SardarPRA2023} and comparable to that of the singlet ground state of CaF+Rb system \cite{SardarCaFRb}. 

Next, we conduct two-parameter geometry optimization for the ground-state geometry of the trimer at the CCSD(T) level, within the rigid-rotor approximation for CaF. The rigid rotor model underestimates the interaction strength at the global minimum of approximately 1572 cm$^{-1}$ for the X$~^2\mathrm{A}'$ state, compared to full-dimensional geometry optimization. However, the shape of the optimized geometry remains the same in both cases.

\begin{table}[!t]
\caption{
Optimized parameters for the global equilibrium geometry of the ground X$~^2\mathrm{A}'$ state of the Ca$_2$F trimer. The equilibrium geometrical parameters $r_\text{CaF}$ and $R$ are given in bohr, $\theta$ in degrees, and the potential well depth $D_e$ in cm$^{-1}$}.
\label{tab:global}
\begin{ruledtabular}
\begin{tabular}{ c c c c c c} 
& $r_\text{CaF}$ & $R$ & $\theta$ & $D_e$ \\
\colrule
Non rigid-rotor   & 4.001 & 5.601&138.38&8973 \\ 
Rigid-rotor   & 3.695 & 5.873&137.49&7401\\ 
\end{tabular}
\end{ruledtabular}
\end{table}

\subsection{1D PESs of trimer}
\label{subsec:1D cuts}
We compute 1D cuts of the PESs for the CaF+Ca interactions in both their ground- and excited-electronic states using the MRCI method. We calculate nine electronic states arising from the lowest three asymptotes of CaF+Ca, considering an excitation energy of approximately 16490 cm$^{-1}$ for CaF. We face the issue of convergence for high-lying electronic states. We use the rigid rotor approximation for CaF. In our calculations, we consider three orientations of CaF+Ca: linear geometries at $\theta = 0^\circ$ and $180^\circ$, and a T-shaped geometry at $\theta = 90^\circ$. For the highly symmetric linear orientations of the trimer, we compute two $^2\Sigma$ states, four $^2\Pi$ states, one $^4\Sigma$ state, and two $^4\Pi$ states, considering the C$_{2\mathrm{v}}$ point group. For the T-shaped configuration, we compute four $^2\mathrm{A}'$ states, two $^2\mathrm{A}''$ states, two $^4\mathrm{A}'$ states, and one $^4\mathrm{A}''$ state under the C$_\mathrm{s}$ point group.

\begin{figure}
\includegraphics[width=\linewidth]{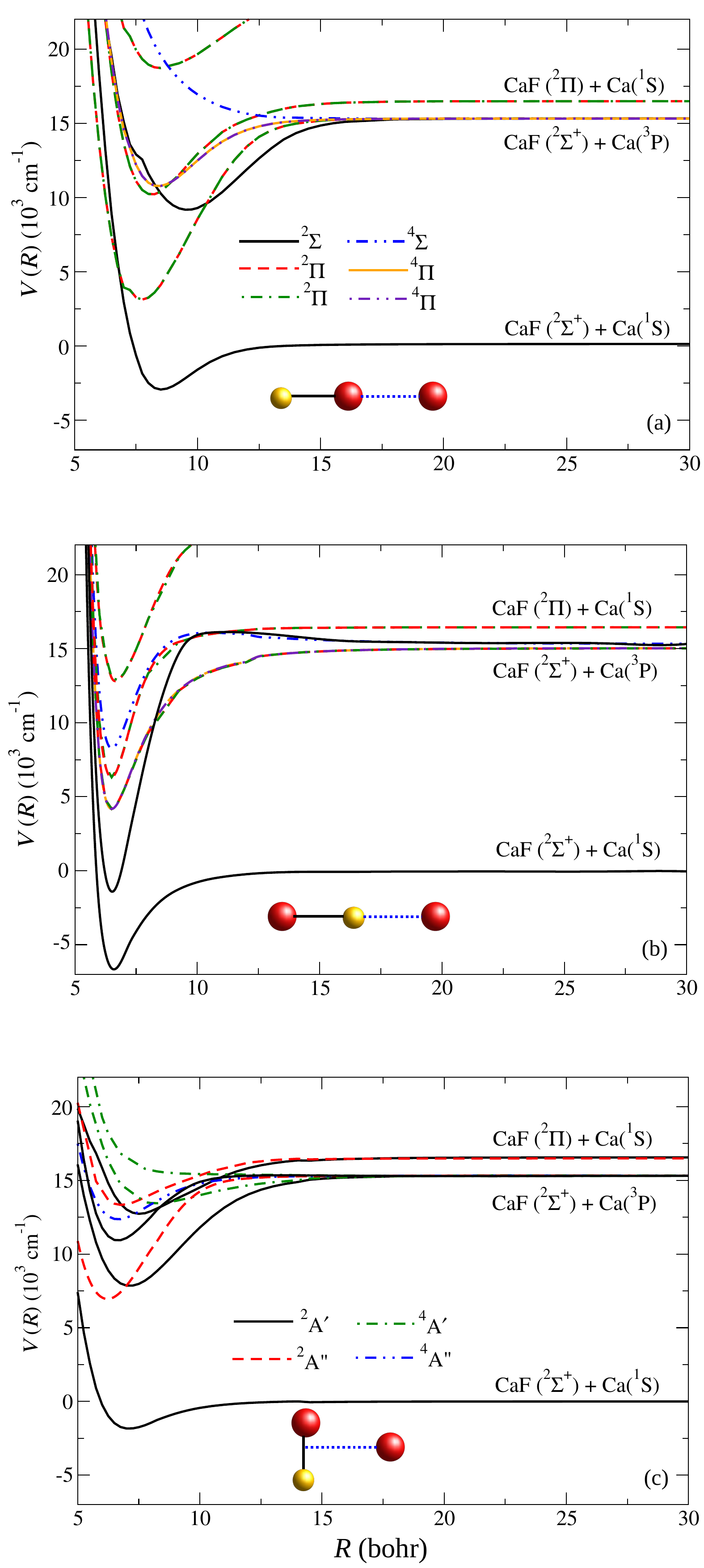}
\caption{\label{fig:1D}One-dimensional cuts of the potential energy surfaces for CaF+Ca as a function of $R$. Panels (a) and (b) show the 1D PESs for the linear configurations at $\theta = 0^\circ$ and $\theta = 180^\circ$ within the C$_{2\mathrm{v}}$ point group, while panel (c) shows the T-shaped configuration under the C$_\mathrm{s}$ point group.}
\end{figure}

In Figure~\ref{fig:1D}, we present the 1D cuts of PES for the Ca$_2$F trimer as a function of the center-of-mass distance $R$. The position of the minimum for the (X)~$^2\mathrm{A}'$ (or X~$^2\Sigma$) state varies with orientations. The PES of the ground state exhibits a significantly deep potential well for the linear $\theta=180^\circ$ orientation. In this configuration, the more electronegative F atom from CaF approaches the second Ca atom, resulting in favorable interactions. In contrast, a shallower potential well is observed for the (X)~$^2\mathrm{A}'$ state in the T-shaped configuration.

\begin{figure*}[t]
\includegraphics[width=\linewidth]{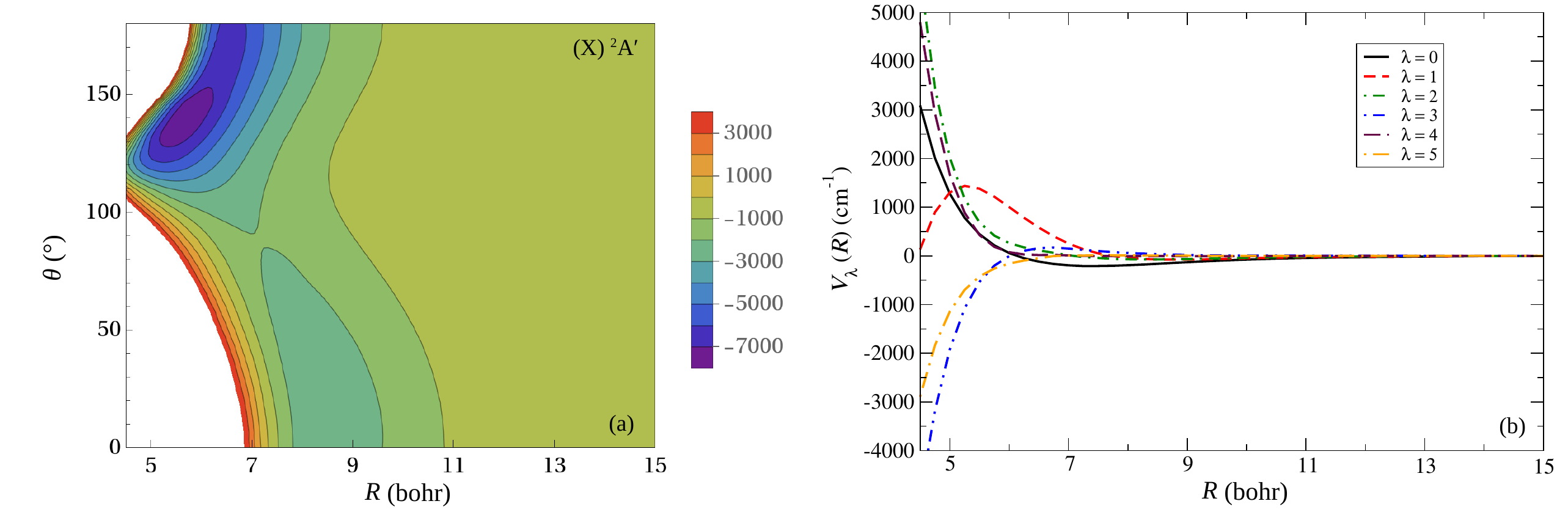}%
\caption{\label{fig:legend} 2D PES and the corresponding Legendre components for the ground (X)~$^2\mathrm{A}'$ state of CaF+Ca in panel (a) and (b). The contour colorbar indicates the interaction energy in cm$^{-1}$. }
\end{figure*}

The excited electronic states of the Ca$_2$F trimer are strongly bound compared to the ground state. In particular, the first excited electronic state (2)~$^2\mathrm{A}'$ or (2)~$^2\Sigma$, is significantly deeper for the linear Ca-F-Ca configuration ($\theta=180^\circ$) compared to other excited states, originating from the CaF($^2\Sigma^+$)+Ca($^3\mathrm{P}$) asymptote. However, the potential well depth for (2)~$^2\mathrm{A}'$ state becomes comparable to or smaller than that of other excited states in different configurations. Notably, for the linear $\theta=  180^\circ$ orientation of CaF+Ca, the (2)~$^2\Sigma$ state approaches the ground-state PES and eventually crosses the ground-state asymptote, CaF($^2\Sigma^+$)+Ca($^1\mathrm{S}$). In addition, we observe a kink in the PES of (2)~$^2\Sigma$ state for the alternative linear geometry F-Ca-Ca ($\theta=0^\circ$). This kink could be attributed to a conical intersection with a $^2\Sigma$ state arising from the higher asymptote. We further analyze this particular state in the following subsection to analyze the chemical reaction pathway between the ground $^2\Sigma^+$ state of CaF and the metastable Ca($^3\mathrm{P}$). 

The PESs for the excited $^2\Pi$ states are significantly deep. We observe a double-well feature for (1)~$^2\Pi$ state in F-Ca-Ca configuration. For linear geometries, each pair of excited $^2\Pi$ states, arising from the CaF($^2\Sigma^+$)+Ca($^3\mathrm{P}$) and CaF($^2\Pi$)+Ca($^1\mathrm{S}$) asymptotes, is energetically degenerate. This degeneracy results from the presence of two degenerate $\Pi$ components, $\mathrm{B}_1$ and $\mathrm{B}_2$, in the C$_{2\mathrm{v}}$ point group. A similar conclusion applies to the degenerate $^4\Pi$ states. However, this degeneracy is lifted in the lower symmetry T-shaped configuration of CaF+Ca.

\subsection{2D PES of trimer}
\label{subsec:2DPES}
We compute 2D PES for the ground (X)~$^2\mathrm{A}'$ state of Ca$_2$F trimer in the Jacobi coordinates. We employ the CCSD(T) method within the rigid-rotor approximation applied to CaF. In Figure~\ref{fig:legend}, we present a 2D contour plot for the ground electronic state and the corresponding Legendre components. The 2D PES and Legendre components are useful for subsequent coupled-channel scattering calculations. The 2D PES exhibits a global minimum and a local minimum. The global minimum appears in a bent geometry, and geometrical parameters, including potential well depth, are exactly similar to those of the optimized geometry for (X)~$^2\mathrm{A}'$ state. The local minimum is located in the F-Ca-Ca configuration ($\theta = 0^\circ$, $R = 8.501$ bohr). These geometries are similar to those of the analogous system, Sr$_2$F \cite{KosickiJPCA2017}. Notably, in the Ca$_2$F trimer, global and local minima are separated by an energy barrier that lies below the ground state asymptote CaF($^2\Sigma^+$)+Ca($^1\mathrm{S}$), suggesting these minima are accessible at ultracold collision energies. Furthermore, pronounced variation in energy on the 2D PES underscores the strong anisotropy of the surface.

In Figure~\ref{fig:legend}(b), we show the radial dependence of the lowest five Legendre components $V_\lambda(R)$, for the 2D PES of the ground (X)~$^2\mathrm{A}'$ state. These Legendre components provide crucial insights into the anisotropy of CaF+Ca interactions. The isotropic component, $V_0 (R)$, exhibits attractive behavior in the short range of $R$ and has a well-defined minimum. In contrast, the first anisotropic term, $V_1 (R)$, is repulsive and its magnitude increases at the short-range molecule-atom separation, contributing a dominant anisotropy to the PES. Higher-order anisotropic terms ($\lambda \geq 2$) also exhibit repulsive behavior, characteristic of short-range interactions.

We determine leading long-range induction and dispersion coefficients for the ground-state interactions between CaF and Ca. These coefficients are useful to study ultracold collisions in CaF+Ca system. We compute $C_{6,0}$ and $C_{6,2}$, both of which are the sum of the corresponding induction and dispersion interactions. $C_{6,0}$ and $C_{6,2}$ are determined by solving Eqs.~\ref{eq:ind}-\ref{eq:disp}. In our calculations, we use the experimentally measured electric dipole moment of CaF and the theoretically recommended static polarizability of Ca. We obtain $C_{6,0}= 1778$ E$_\mathrm{h}a_0^6$ and $C_{6,2}=95$ E$_\mathrm{h}a_0^6$, where E$_h$ is the Hartree energy and $a_0$ is the Bohr radius. The computed $C_{6,0}$ value for CaF+Ca is approximately 48\% lower than SrF+Sr\cite{KosickiJPCA2017} and 75\% lower than SrOH+Sr \cite{Kosicki2025arxiv} systems, which are obtained by the curve fitting method. The lower $C_{6,0}$ value for CaF+Ca indicates that this system has weak dispersion interactions compared to Sr-containing systems, which could be attributed to the lower polarizability of Ca \cite{SchwerdtfegerMP2019}. 

\subsection{Chemical reactions}
\subsubsection{Ground \textsuperscript{1}S state of Ca}
\label{subsec:long-range}
\begin{figure}[b]
\includegraphics[width=0.45\textwidth]{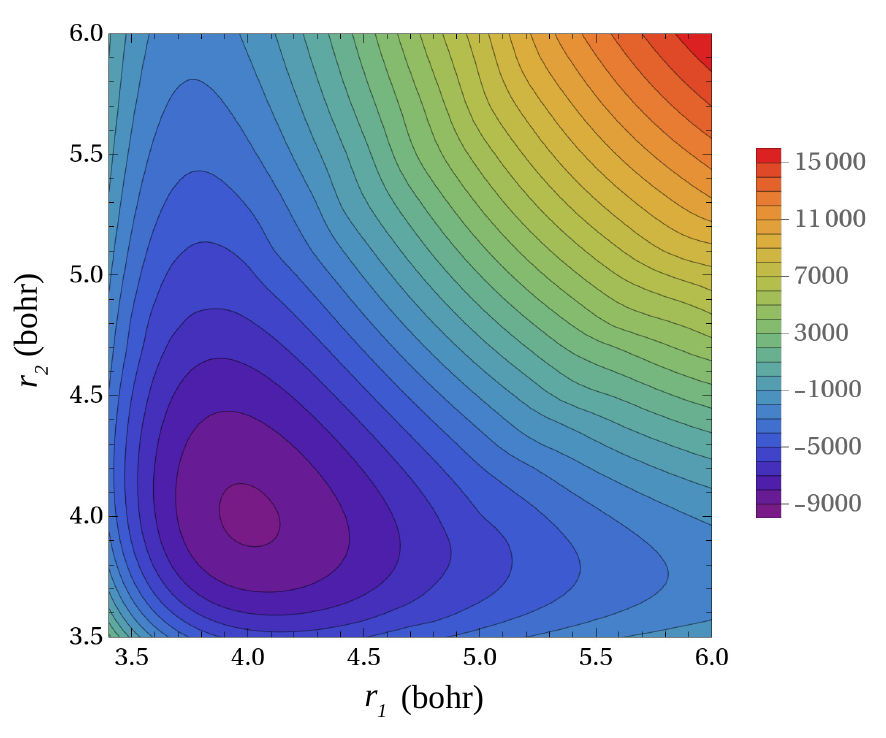}%
\caption{\label{fig:gsrxn} 2D PES for atom exchange in the ground (X)~$^2\mathrm{A}'$ state of CaF+Ca. The bond angle Ca-F-Ca is fixed at the global equilibrium geometry. The interaction energy (in cm$^{-1}$) is calculated with respect to the ground-state asymptote.  }
\end{figure}
We study chemical reactions in the CaF+Ca system to understand possible reaction pathways. The direct chemical reaction CaF+Ca $\rightarrow$ Ca$_2$ + F is highly endothermic and energetically forbidden, attributed to the stronger binding energy of CaF compared to Ca$_2$. Therefore, we focus on the isotope-exchange reactions between CaF and Ca. 

We first investigate the reaction pathway for the ground-state interaction between CaF and Ca. Using the CCSD(T) method, we compute the PES in internal Z-matrix coordinates with the representation Ca-F-Ca. In this coordinate, one of the Ca-F bond distances is denoted by $r_1$, another Ca-F bond length is represented by $r_2$, and Ca-F-Ca bond angle is fixed at the global equilibrium minimum of the trimer. We vary the interatomic distances $r_1$ and $r_2$ from 3.5 to 6 bohr, approximately twice the bond length of the CaF monomer. The contour maps of the PES for the separation of fluorine and calcium atoms are presented in Figure~\ref{fig:gsrxn}.

The PES for the trimer exhibits smooth behavior with a well-defined single minimum, indicating a stable configuration during the interaction. The geometric parameters at this minimum closely match those obtained from full-dimensional optimization, including the equilibrium bond lengths and the depth of the potential well. Notably, no energy barriers are observed along the reaction coordinates for the relevant geometries, suggesting that the atom-exchange process between CaF and Ca occurs without hindrance in their ground electronic states. This barrierless behavior indicates that the reaction can proceed efficiently at low collision energies, allowing exchange of the calcium atom between the fluorine-bound and free calcium atom. Such characteristics are consistent with a barrierless reaction pathway in SrF+Sr \cite{KosickiJPCA2017,KosickiJPB2020}, where the rearrangement of atomic positions is energetically favorable, allowing the system to evolve smoothly from reactants to products. The absence of intermediate transition states further supports the feasibility of the atom-exchange reaction as a likely pathway in the ground-state interactions of CaF with Ca.

Such a ground-state atom-exchange reaction can be controlled by an appropriate choice of the calcium isotope. Here, we consider the various thresholds for different isotopes of Ca in reactions like
\begin{equation}
  ^\mathrm{A}{\mathrm{CaF}} + ^\mathrm{B}{\mathrm{Ca}}\rightarrow ^\mathrm{B}{\mathrm{CaF}} + ^\mathrm{A}{\mathrm{Ca}}.
  \label{eq:isotope}
\end{equation}
We assume that the energy of an atomic isotope, $^\mathrm{A}$Ca, is determined purely by the electronic structure and is independent of the isotope. For the CaF molecule, the Born-Oppenheimer potential $V_{\mathrm{CaF}}(r)$ between a Ca atom and an F atom is likewise determined by \textit{ab initio}  electronic structure methods and does not depend on isotope. Consequently, isotope effects arise only through the reduced mass, $\mu$, of the CaF molecule, which influences the equilibrium vibrational frequency:
\begin{equation}
    \omega_e=\sqrt{\frac{1}{\mu}\frac{d^2V_{\mathrm{CaF}}}{dr^2}}\Big|_{r_e}.
\end{equation}
The zero-point energy (ZPE) of this CaF potential, which sets the relative scattering thresholds for different isotopes, is therefore 
\begin{equation}
\mathrm{ZPE} = \frac{1}{2}\hbar \omega_e=\frac{1}{2}\hbar \sqrt{\frac{1}{\mu}\frac{d^2V_{\mathrm{CaF}}}{dr^2}}\mid_{r_e}.    
\end{equation}
 Here, $\mu$ is the reduced mass of the Ca-F system and ${r_e}$ is the equilibrium position of the CaF molecule.

\begin{table}[b]
\begin{ruledtabular}
\caption{\label{tab:isotope} Energy change (in cm$^{-1}$) for the isotope-exchange reactions between $^\mathrm{A}$CaF and $^\mathrm{B}$Ca, where A and B are the different isotopes of calcium atom.   }
\begin{tabular}{ c c c c c c c c} 
&$^{40}\text{CaF}$ & $^{42}\text{CaF}$& $^{43}\text{CaF}$  & $^{44}\text{CaF}$ & $^{46}\text{CaF}$ & $^{48}\text{CaF}$  \\
\colrule
$^{40}\text{Ca}$   & 0      & 2.247& 3.301&4.305 &6.199 &7.947 \\ 
$^{42}\text{Ca}$   & -2.247 & 0    &1.053 &2.059  & 3.953&5.701 \\ 
$^{43}\text{Ca}$   & -3.301 & -1.053&0    &1.006 & 2.899 &4.647\\ 
$^{44}\text{Ca}$   & -4.305 & -2.059&-1.006&0    & 1.894 & 3.642\\ 
$^{46}\text{Ca}$   & -6.199 & -3.953&-2.899&-1.894 & 0 & 1.748\\ 
$^{48}\text{Ca}$   & -7.947 & -5.701&-4.647&-3.642 & -1.748 & 0\\ 
\end{tabular}
\end{ruledtabular}
\end{table}

In Table~\ref{tab:isotope}, we present the energy change in reaction~(\ref{eq:isotope}) for six different isotopes of Ca. In particular, the amount of energy change in reaction~(\ref{eq:isotope}) is computed by subtracting the ZPE of the product $^\mathrm{B}$CaF from the reactant $^\mathrm{A}$CaF. We note that the reaction (\ref{eq:isotope}) is exothermic only if the calcium isotope in the product CaF molecule has a greater mass than that in the reactant. For example, the reaction $^{40}$CaF + $^{42}$Ca $\rightarrow$ $^{42}$CaF + $^{40}$Ca is energetically allowed, while the reverse reaction, $^{42}$CaF + $^{40}$Ca $\mathrel{\not\rightarrow}$ $^{40}$CaF + $^{42}$Ca, is endothermic and therefore energetically forbidden. 

The amount of energy released in the above exothermic isotope-exchange reactions lies in the range 1–8 cm$^{-1}$. This energy is two orders of magnitude smaller than the equilibrium vibrational spacing $\omega_e = 581$ cm$^{-1}$ of $^{40}$CaF \cite{KramidaNIST2018}, but an order of magnitude larger than the first rotational excitation energy, $2B_e$ (with $B_e = 0.338$ cm$^{-1}$ for $^{40}$CaF \cite{KramidaNIST2018}). Consequently, the product $^\mathrm{B}$CaF can remain in its vibrational ground state while being rotationally excited in isotope-exchange reactions involving $^\mathrm{A}$CaF and $^\mathrm{B}$Ca. Similar isotope-exchange reactions have been studied for heteronuclear dimers, where the product molecules occupy in their absolute rovibrational ground states \cite{TomzaPRL2015}.
The situation where the products of reaction can be rotationally, but not vibrationally, excited is analogous to what was observed in KRb + KRb ultracold reactions \cite{LiuNature2021}.  In that case, detection of all possible rotational final states allowed the experiment to test statistical models of the reaction.  An analogous opportunity presents itself in CaF+Ca reactions, with the added experimental knob of isotope selection.

As part of this analysis, it will matter whether the exit channels are in the Winger threshold regime.
 To this end, we compute the characteristic van der Waals length scale $R^*$ and related energy scale $E^*$ expressed as:
\begin{equation}
  R^* = \Bigr( \frac{2\mu C_{6,0}}{\hbar^2}\Bigl)^{1/4} \hspace{0.2cm}\textrm{and}\hspace{1.0cm} 
  E^* = \frac{\hbar^2}{2\mu (R^*)^2},
  \label{eq:lengthS}
\end{equation}
where $\mu$ is the reduced mass for $^{40}$CaF+$^{40}$Ca and $\hbar$ is the reduced Planck constant. We determine the van der Waals length scale $R^*=111.6~a_0$, and energy scale $E^*=292$ microkelvin. The computed energy scale for CaF+Ca is slightly higher than that of the alkali-metal dimers. However, in the isotope exchange chemical reaction between CaF and Ca, the exothermic channel lies in a few Kelvin range, much larger than $E^*$. Thus, the exothermic channel occurs well above the Wigner threshold regime.

\subsubsection{Metastable \textsuperscript{3}P state of Ca}

\begin{figure}[t]
\includegraphics[width=0.45\textwidth]{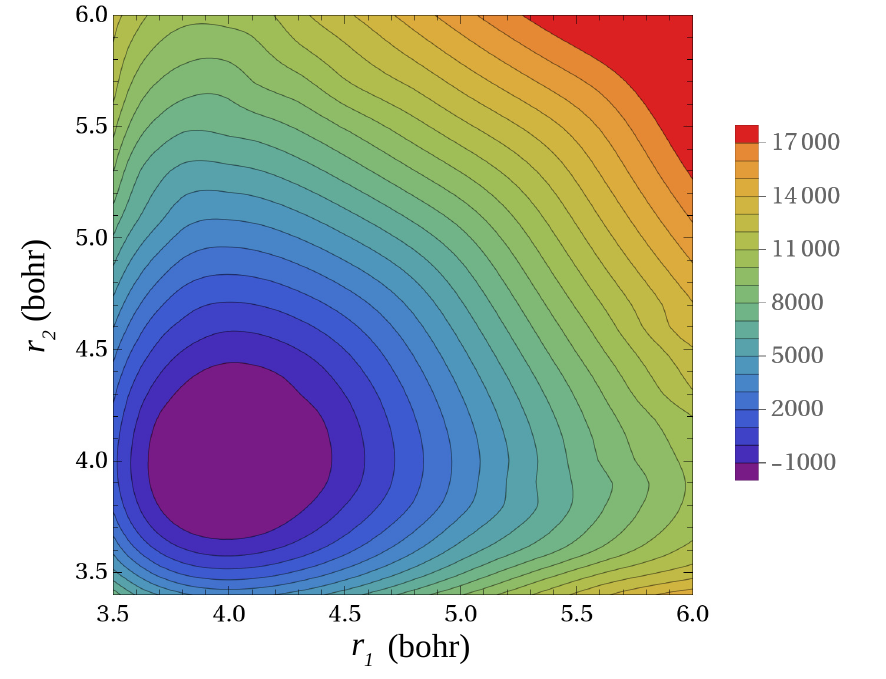}%
\caption{\label{fig:exrxn} 2D PES for the (2)~$^2\Sigma$ state of CaF+Ca. Energies are referenced to the ground-state asymptote of the Ca$_2$F trimer and expressed in cm$^{-1}$. }
\end{figure}

We further investigate the atom-exchange reaction pathway between the ground state of CaF and the excited state of Ca. The excited $^3$P state of Ca is a metastable state \cite{FurcinittiPLA1975} and could be experimentally relevant to study excited state collision dynamics in CaF+Ca. The chemical reaction in the incident channel,  CaF$(^2\Sigma^+)$+Ca$(^{3}\mathrm{P})$, can proceed through several pathways: 1) non-adiabatic transitions driven by conical intersections or avoided crossings between the PESs of the  (X)~$^2\mathrm{A}'$ and (2)~$^2\mathrm{A}'$ states; 2) a non-adiabatic transition between the  (2)~$^2\mathrm{A}'$ and (X)~$^2\mathrm{A}'$ states induced by strong spin-orbit mediated coupling, arising from the presence of the metastable $^3$P state of Ca; or 3) a spontaneous radiative pathway with emission of a photon.

To explore the atom-exchange pathway in the CaF($^2\Sigma$)+Ca($^3\mathrm{P}$) incident channel, we compute the PES for the excited (2)~$^2\Sigma$ state by the MRCI method, using the same Z-matrix coordinates with the Ca-F-Ca angle fixed at $180^\circ$. From the 1D cut of PES as shown in Figure~\ref{fig:1D}, we observe that the (2)~$^2\Sigma$ state crosses the ground state asymptote; the Ca–F–Ca angle is maintained at $180^\circ$ in subsequent 2D PES calculations. The resulting 2D PES for the (2)~$^2\Sigma$ state is shown in Figure~\ref{fig:exrxn}. The interaction energy is computed with respect to the ground state asymptote.

The surface shown in figure~\ref{fig:exrxn} crosses into negative values, indicating that the excited (2)~$^2\Sigma$ state (concerning the incident CaF($^2\Sigma$)+Ca($^3\mathrm{P}$) channel) crosses the ground-state CaF($^2\Sigma$)+Ca($^1\mathrm{S}$) asymptote of the trimer. This state lies more than 1000 cm$^{-1}$ below the ground-state asymptote, indicating the possibility of mixing between the rovibronic levels of the ground (X)~$^2\Sigma$ and excited (2)~$^2\Sigma$ states, as well as mixing between the rovibronic states of the (2)~$^2\Sigma$ manifold and the ground-state scattering continuum. Such rovibronic mixing driven by spin-orbit coupling could be strong, due to the presence of a Ca atom in the excited $^3\mathrm{P}$ state. This may result in a nonadiabatic electronic transition from the excited (2)~$^2\Sigma$ electronic state to the ground state, thus opening a pathway for chemical reactivity. However, in the present study, we do not check whether such a reaction is barrierless or not.

Furthermore, our present calculations for extensive scans of the PESs across different geometries and symmetries revealed no direct crossings between the PESs associated with the entrance and exit reaction channels. Notably, the PES for the entrance channel intersects with high-lying electronic states. But, we did not identify any viable pathway connecting the entrance and exit channels through surface crossings of different symmetries, like $^2\Sigma$-$^2\Pi$ in linear geometries and $^2\mathrm{A}'$-$^2\mathrm{A''}$ in non-linear geometries.

\section{Conclusions}
\label{sec:conclusion}
We have investigated ultracold  CaF and Ca interactions in their ground and excited electronic states. Using state-of-the-art quantum chemistry methods, we computed nine electronic states that arise from the three lowest asymptotes of CaF+Ca. The global equilibrium minimum for the electronic ground (X)~$^2\mathrm{A}'$ state of trimer is symmetric and bent with a potential well depth of approximately 8973 cm$^{-1}$. Under the rigid rotor approximation, 2D PES for the (X)~$^2\mathrm{A}'$ state exhibit highly anisotropic behavior in short-range molecule-atom separations. The excited states of the Ca$_2$F trimer are strongly bound. Notably, the PES for the first excited (2)~$^2\mathrm{A}'$ state concerned with CaF($^2\Sigma$)+Ca($^3\mathrm{P}$) asymptote is substantially deep and eventually crosses the ground state asymptote. 

We studied atom-exchange reaction pathways between ground-state CaF and Ca, as well as with the excited metastable state of Ca. For the ground-state reaction between CaF+Ca, no barriers have been found along the reaction coordinates. This suggests that the reaction can proceed efficiently at ultracold energy regimes, allowing transfer of the Ca atom between the F-bound and free Ca atom. Moreover, this atom-exchange reaction can be controlled through the appropriate choice of a Ca isotope. 

To understand the reaction pathway between CaF($^2\Sigma$) and Ca($^3\mathrm{P}$), we investigated 2D PES for the excited (2)~$^2\mathrm{A}'$ state, under the non-rigid rotor approximation for CaF. We observed that (2)~$^2\mathrm{A}'$ state lies more than 1000 cm$^{-1}$ below the ground-state asymptote, enabling mixing between the rovibronic states of (X)~$^2\mathrm{A}'$ and (2)~$^2\mathrm{A}'$ electronic states. Additional mixing occurs between the rovibronic states of (2)~$^2\mathrm{A}'$ and the ground-state scattering continuum. Such mixing induced by spin-orbit coupling could allow for a non-adiabatic transition from the excited state to the ground electronic state. However, we did not observe any direct crossing between PESs of these two states. 

Finally, while not the focus of the present paper, we note that optically allowed transitions between the ground state and electronically excited states would be a valuable next step in this analysis.  In particular, such transitions afford exciting possibilities such as atom-molecule photoassociation as a means of producing polyatomic ultracold molecules.  Note that even more exotic processes have been contemplated, such as the creation of asymmetric tops, in the ultracold regime \cite{Kosicki2025arxiv}.

Overall, the spectroscopic properties for CaF+Ca, including the atom-exchange reactions and their control, are important in the context of ultracold chemistry. These results motivate further investigations of interactions in both ground and excited electronic states for similar F-containing, directly laser-cooled molecules and their corresponding metal atoms, such as AlF+Al, SrF+Sr, and BaF+Ba.

\begin{acknowledgments}
D.S. gratefully acknowledges the National Science Centre Poland
(grants No.~2020/38/E/ST2/00564) for financial support and the
Poland's high-performance computing infrastructure PLGrid (HPC
Centers: ACK Cyfronet AGH) for providing computer facilities and
support (computational grant No.~PLG/2024/017527).
JLB acknowledges support from the AFOSR Multidisciplinary University Research Initiative Grant No. GG016303.
\end{acknowledgments}

\bibliography{sample.bib}
\end{document}